\documentclass[12pt,dvips]{article}
\usepackage{amscd,verbatim}
\usepackage[all]{xy}
\usepackage{graphicx}

\usepackage{amsmath}
\usepackage[psamsfonts]{amssymb}
\usepackage{amsthm}
\usepackage{euscript}

\usepackage{latexsym}

\renewcommand{\(}{\begin{equation}}
\renewcommand{\)}{end{equation} \vspace{-.05in}\linebreak}

\newcounter{saveeqn}
\newcounter{savealpheqn}

\newcommand{\alpheqn}{\setcounter{saveeqn}{\value{equation}}%
  \stepcounter{saveeqn}\setcounter{equation}{0}%
  \renewcommand{\theequation}{\mbox{\arabic{section}.\arabic{saveeqn}
\alph{equation}}}
  \renewcommand{\)}{\end{equation}}}
\def\part#1{\frac{\partial}{\partial{#1}}}%
\def\group#1{\refstepcounter{equation}\setcounter{saveeqn}{\value{equati 
on}}%
  \label{#1}\setcounter{equation}{0}%
\renewcommand{\theequation}{\mbox{\arabic{section}.\arabic{saveeqn}
\alph{equation}}}
  \renewcommand{\)}{\end{equation}}}
\newcommand{\reseteqn}{\setcounter{equation}{\value{saveeqn}}%
  \renewcommand{\theequation}{\arabic{section}.\arabic{equation}}%
  \renewcommand{\)}{\end{equation}}}

\newcommand{\aalpheqn}{\setcounter{saveeqn}{\value{equation}}%
  \stepcounter{saveeqn}\setcounter{equation}{0}%
  \renewcommand{\theequation}{\mbox{
        \Alph{subsection}.\arabic{saveeqn}\alph{equation}}}
   \renewcommand{\)}{\end{equation}}}
\newcommand{\areseteqn}{\setcounter{equation}{\value{saveeqn}}%
  \renewcommand{\theequation}{\Alph{subsection}.\arabic{equation}}%
  \renewcommand{\)}{\end{equation}}}

\renewcommand{\thefootnote}{\alph{footnote}}
\renewcommand{\(}{\begin{equation}}
\renewcommand{\)}{\end{equation}}
\newcommand{\ba}{\begin{eqnarray}}
\newcommand{\ea}{\end{eqnarray}}

\newcommand{\bp}{\mathop{\vtop{\ialign{##\crcr
   $\hfil\displaystyle{}\hfil$\crcr\noalign{\kern-13pt\nointerlineskip}
   \BIG{(}\hskip0pt\crcr\noalign{\kern3pt}}}}}
\newcommand{\cbp}{\mathop{\vtop{\ialign{##\crcr
   $\hfil\displaystyle{}\hfil$\crcr\noalign{\kern-13pt\nointerlineskip}
   \BIG{)}\hskip0pt\crcr\noalign{\kern3pt}}}}}
\newcommand{\pa}{\mathop{\vtop{\ialign{##\crcr
    
$\hfil\displaystyle{\oplus}\hfil$\crcr\noalign{\kern+1pt\nointerlineskip 
}
   \hspace{.08in}$^{\alpha=0}$\hskip6pt\crcr\noalign{\kern3pt}}}}}

\newcommand{\R}{\ensuremath{\mathbb R}}

\newcommand{\cF}{\ensuremath{\mathcal F}}

\newcommand{\C}{\ensuremath{\mathbb C}}

\newcommand{\Z}{\ensuremath{\mathbb Z}}

\newcommand{\beq}{\begin{equation}}
\newcommand{\eeq}{\end{equation}}

\numberwithin{equation}{section}

\catcode`\@=11
\def\vereq#1#2{\lower3pt\vbox{\baselineskip1.5pt \lineskip1.5pt
\ialign{$\m@th#1\hfill##\hfil$\crcr#2\crcr\sim\crcr}}}
\catcode`\@=12

\makeatletter
\newcommand\figcaption{\def\@captype{figure}\caption}
\newcommand\tabcaption{\def\@captype{table}\caption}
\makeatother
\renewcommand{\(}{\begin{equation}}
\renewcommand{\)}{\end{equation}}

\newcommand{\ZZ}{{\mathbb Z}}
\newcommand{\QQ}{{\mathbb Q}}

\theoremstyle{plain}

\theoremstyle{definition}

\begin{document}

\begin{titlepage}
\begin{flushright}

hep-th/0701232
\end{flushright}

\vspace{2em}
\def\thefootnote{\fnsymbol{footnote}}

\begin{center}
{\Large\bf  A Higher Twist in String Theory}
\end{center}
\vspace{1em}

\begin{center}
\Large Hisham Sati \footnote{E-mail:
hisham.sati@yale.edu}
\end{center}

\begin{center}
\vspace{1em}
{\em { Department of Mathematics\\
Yale University\\
New Haven, CT 06520\\
USA}}\\
\end{center}

\vspace{0.5cm}
\begin{abstract}
\noindent
 Considering the gauge field and its dual in heterotic string theory as a unified 
field, we show that the equations of motion at the rational level contain a twisted differential with a novel degree seven twist. This generalizes the usual degree three twist that lifts to twisted K-theory and raises the natural question of whether at the integral level the abelianized gauge fields belong to a generalized cohomology theory.  Some remarks on possible such extension are given.

\end{abstract}

\vfill

\end{titlepage}
\setcounter{footnote}{0}
\renewcommand{\thefootnote}{\arabic{footnote}}

\pagebreak
\renewcommand{\thepage}{\arabic{page}}

\section{Introduction}   

The classification of the Ramond-Ramond fields in type II string
theory has been an important piece of progress in recent years.
In the absence of background fields, those are classified by
K-theory of spacetime \cite{MW, FH}. In the presence of the Neveu-Schwarz fields,
the RR fields are then described by twisted K-theory. The twisting in type II string theory
 comes from the NSNS sector. Of
particular interest is the twist coming from the rank three field
$H_3$ which shows up in IIA and IIB string theory. 
  One can pass from (twisted) K-theory to (twisted) cohomology through the
(twisted) Chern character, which is considered as a map from the former to
the latter. In general, what is detected by (twisted) K-theory that is
different from that of (twisted) cohomology is the torsion  
information. At the rational level, the two descriptions coincide, and
cohomology is isomorphic to K-theory. Then, on the cohomology side, at the 
rational level, the fields of classical 
supergravity satisfy the equations that specify de Rham cohomology, namely
$dF=0$ with the nilpotency condition $d^2=0$.
Similar description for type I string theory can be given in terms of twisted
$KO$ theory ( see \cite{MMSt}). What about heterotic string theory?

\vspace{3mm}
In the supergravity multiplet of heterotic string theory there is only one potential $B_2$,
whose field strength is $H_3$. The natural question is whether there is a generalized 
cohomology description of this in analogy to what happens in type I and type II
string theories. 
Freed \cite{Fr} classified $H_3$ and its dual $H_7$, with potentials $B_2$ and $B_6$, 
via $KOSp$-theory. The charges associated 
with $B_2$ lie in $KO^0(X)$, but due to the presence of a magnetic current with 
a self-duality condition, the field $B_2$ itself does not belong to 
$KO^1(X)$ but to the cohomology theory 
$KOSp^{\bullet}=KO^{\bullet} \times KSp^{\bullet}$.

\vspace{3mm}
We know that in heterotic theories there is a coupling between the $H$-field
and the Chern-Simons form of the gauge theory via the Manton-Chapline 
coupling \cite{CM, BddN}. This suggests that the gauge fields, being related to $H$ that way,
might have an interpretation of their own in terms of generalized cohomology. 
It is the purpose of this note to uncover such a structure. We work at the rational
level and then propose the generalized cohomology lift. 
We are thus looking at a cohomology theory related to the 
gauge fields, and while reference \cite{Fr} looks at Maxwell's system for $H_3$ and $H_7$, 
we consider the system for $F_2$ and $*F_2$.

\vspace{3mm}
 Note that a curvature being a K-theory element already appears in
type IIA string theory where $F_2=dA$ (in the constant dilaton case,
see \cite{MS}) which is the curvature of the spacetime bundle-- the
M-theory circle bundle -- and is
interpreted as the RR 2-form, and fits into the K-theoretic description
of the total RR field. We hope that, with this in mind, the transition to 
considering the gauge fields in heterotic string theory as elements in generalized 
cohomology should perhaps not sound too conceptually strange. 

\vspace{3mm}
We assume an abelian reduction of the Yang-Mills group. 
We define a total gauge field which satisfies a twisted Bianchi identity, except that
now the twist is given by the degree seven field $H_7$. We consider the corresponding
complex and the generator leading to a uniform degree differential. The appearance
of a higher degree generator, which we identify, makes contact with the discussion 
on higher generalized cohomology in string theory \cite{KS1, KS2, KS3, S4} and 
in M-theory \cite{S1, S2, S3}.

\vspace{3mm}
The note is organized as follows. After reviewing the standard 3-form twist, given by 
the NS $H$-field, on the Ramond-Ramond fields, we present a degree seven twist
in heterotic string theory. What is being twisted is not the RR fields, but rather the
abelianized {\it gauge} fields. We show that the twist makes up a differential, i.e.
the expression squares to zero. As usual, it is tempting to lift the rational equation
to include torsion.  We do not have a final answer on which generalized cohomology 
theory is the right one, but we discuss some possibilities which fit into the discussion 
of the RR fields for type II string theory \cite{KS1, KS2, KS3, S4} and for type I \cite{Fr}. 
We should point out a caveat. This is  
that we do not have 
a clear handle on heterotic string theory like we do on type II string theory. In the latter
case, a number of points support the proposal of a relationship between K-theory
and string theory. Among the most obvious are
the observation about RR charges and K-theory \cite{MM} and 
the relationship between the geometry of K-theory and the 
geometry of Chan-Paton bundles over D-branes.
However, the situation in heterotic string theory is far from being analogous. 
The arguments provided here are based on duality.  
The final answer could be arrived at through 
a derivation from the structure of the theory, or through an example which includes 
subtle torsion fields. Both are outside the scope of this note.

\section{Review of the Three-Form Twist in Type II}
Here we recall the known case.
In string theory, motivated by K-theory, one can combine the Ramond-Ramond
of different ranks, into one RR object as
\(
F=\sum_{i}^{n}F_i,
\)
where $i=2p$ for IIA ($i=2p+1$ for IIB) and $n=10$ for IIA 
(and $n=9$ for IIB).
Here we have a twist by the B-field
\(
d_{H_3}=e^{B_2}de^{-B_2}=(d-dB_2 \wedge)=(d-H_3\wedge).
\label{H3}
\)
This differential satisfies
\(
d_{H_3}^2=d^2 + H_3 \wedge d - H_3 \wedge d - dH_3 \wedge -H_3 \wedge H_3 \wedge,
\)
which gives zero because of (\ref{H3}),
and the fact that the wedge product of two copies of an abelian 
odd-degree form is zero. Thus it defines a twisted form of de Rham theory
$\Omega^k(X)$ but with the differential $d_{H_3}$.
This can be 
lifted to twisted K-theory.
Although the above presentation involved
a cohomologically trivial $H$-field, the result is of course the same for
the case $[H_3]\neq 0$.
The equation of motion of the RR fields in twisted cohomology is 
\(
dF_n - H_3 \wedge F_{n-2}=0.
\)

\section{The Seven-Form Twist in the Heterotic Theory}
Let us start with a general action of the form
\(
S=\int H_3 \wedge * H_3 + \int F_2 \wedge * F_2
\)
with a Chapline-Manton coupling $H_3=CS_3(A)$,
where $CS_3(A)$ is the Chern-Simons three-form for the gauge field
$A$, whose curvature is $F_2=d_A A$ ($d_A$ is the gauge covariant 
derivative). This is part of the coupling of type I supergravity to 
Yang-Mills theory, say in heterotic string theory. 
We would like to consider the case where the Yang-Mills group, 
$E_8 \times E_8$ or ${\rm Spin}(32)/\Z_2$, is broken down to an abelian
subgroup, We assume manifolds $M^{10}$ such that this breaking via Wilson
lines is possible. We could consider the Cartan torus for example. 

\vspace{3mm}
Let us vary the action with respect to $A$ in order to get
the gauge field equation of motion. We have (assuming the abelian 
case)
\(
\frac{\delta S}{\delta A} = * H_3 \wedge F_2 - d* F_2 
\)
which implies the equation for the gauge field,

\(
(d* - * H_3 \wedge) F_2=0 .
\)
We manipulate the above equation to put it in 
a more suggestive form. Namely, in order to write the operator
as $(d - \mathcal{O})$, we define the combined curvature
\(
{\mathcal F}=F_2 + * F_2 
\)
of the gauge field strength and its dual. The gauge field equation
then will be written as
\(
(d - * H_3 \wedge) {\mathcal F}=0.
\)
This suggests that the combined object  ${\mathcal F}$ is the analog of the situation
in the case of the Ramond-Ramond fields in type II string theories. We can apply 
the ten-dimensional Hodge ``$*$" operator to $H_3$ to 
give $H_7$, and then the equation would be written in an even nicer form
\(
(d - H_7\wedge){\mathcal F}=0 .
\)
We can check whether the new operator is actually a differential 
in a complex -- a priori some higher-twisted de Rham complex. Then we
can check whether this can be argued to come from higher twisted K-theory
or other higher twisted generalized cohomology theory.  

\vspace{3mm}
Let us first check whether there is a complex. Let us compute the 
square of the shifted differential
\(
(d-H_7)^2 = d^2 + H_7 \wedge d  - H_7 \wedge d - dH_7 \wedge + H_7 \wedge H_7 \wedge,
\)
which is zero for reasons analogous to those in the case of the three-form twist, namely  
because $H_7$ is a closed odd form.
In the original 
Green-Schwarz formulation \cite{GS} of the heterotic anomalies, one takes 
$dH_7=0$. Therefore, in this situation we can define the twisted 
differential 
\(
d_{H_7}=d-H_7
\)
which squares to zero, $d_{H_7}^2=0$.

\section{ Possible Lift to Generalized Cohomology}
What we seek is a generalized cohomology theory $Z$ such that
\begin{enumerate}
\item It is possible to twist $Z$-theory by $H_7$. 
\item The ``Chern character" for $H_7$-twisted $Z$-theory takes values in 
$H_7$-twisted de Rham cohomology. 
\item The class ${\mathcal{F}}$ is the image under this twisted Chern character 
of a more fundamental class ${\mathcal{G}}$ in $H_7$-twisted $Z$-theory. 
\item The differential $d_7$ in the Atiyah-Hirzebruch spectral sequence 
for $Z$-theory is a topological obstruction in heterotic string theory. 
\end{enumerate}
Looking for a cohomology theory $Z$ with a particular twist is a little like looking for 
a ring $R$ whose group of invertible elements $R^{\times}$ 
contains a particular subgroup. 
For example, the field $\C$ has units $\C^{\times}$. The group-ring 
$\Z[\C^{\times}]$ also has $\C^{\times}$ as its group of units. Thus, starting with
the group of units we cannot expect to pin down the desired theory without 
any further constraints. Unfortunately, however, little is available about
the relationship between the geometry of heterotic string theory and the 
geometry of the desired generalized cohomology theory, the reason being
partly due to the lack of knowledge about the geometry associated to 
generalized cohomology theories (beyond K-theory, that is ) in general. 
We will nevertheless proceed with some plausibility arguments and 
consider candidates for such a theory.

\vspace{3mm}
 Recall the case of twist in K-theory \cite{Tel} \cite{AS}: A twisting of complex
K-theory over $M$ is a principal $BU_{\otimes}$-bundle over $M$.
This can then be factorized into
$BU_{\otimes} \equiv K(\ZZ, 2) \times BSU_{\otimes}$.
Then the twisting is a pair $\tau=(\delta, \chi)$ consisting of a
determinantal twisting $\delta$, which is a $K(\ZZ,2)$-bundle over
$M$ and a higher twisting $\chi$, which is a $BSU_{\otimes}$-torsor.
Twistings are classified, up to isomorphism, by a pair of classes
$[\delta]\in H^3(M,\ZZ)$ and $[\chi]$ in the generalized cohomology group
$H^1(X,BSU_{\otimes})$.
The former is what is usually referred to as twisted K-theory \cite{Ros}
\cite{BCMMS}, where the twist is given by the Dixmier-Douady
class. The latter is of interest to us. The twistings of the 
rational K-theory of $X$
are classified, up to isomorphism, by the group \cite{Tel}
$\prod_{n>1} H^{2n+1} (M;\QQ)$.
This shows that, in addition to the usual $H^3(M)$-twisting, one
can in principle have twistings from $H^5(M)$ and $H^7(M)$ and so on.

\vspace{3mm}
Is this what we are after? If this is the correct interpretation then we should be able to 
form a twisted complex using $H_7$ in the context of K-theory. One immediate problem 
is that the Bott generator in K-theory does not have the right dimension to be used in
forming a complex. Another problem is that including any twist beyond dimension three
of \emph{integral} K-theory would force us to include all higher twists as well.
\footnote{We thank Constantin Teleman for explaining this point.}
Thus we will look beyond K-theory. However, the above argument is not totally lost 
as one might be able to find a twisted generalized cohomology theory whose twists include 
those of K-theory.
\footnote{We thank Matthew Ando for earlier explanation on this point.}
At least in the case of the determinantal twist, it was shown in \cite{Doug} that 
twists of K-theory are included as ``elliptic line bundles" 
in the theory of topological modular forms,
a fact that was used in \cite{KS2}. There is a map $K(\Z,3)\to TMF$ coming from the String 
orientation. A twist of K-theory of the form $X \to K(\Z,3)$ 
gives rise to an  element of $TMF(X)$ by composition $X \to K(\Z,3) \to TMF$.  
The map $X \to K(\Z, 3)$ corresponds to a $BS^1$ bundle over $X$ which can
be described geometrically as a 1-dimensional 2-vector bundle, hence a
higher notion of a line bundle. 

\vspace{3mm}
Here we present two possibilities. We start with the first one.
We use the combination 
\(
\mathcal{F}=u_1^{-1} F_2 + u_2^{-1}F_8,
\)
where $u_1$ has degree two and $u_2$ has degree eight. 
The above expression has a uniform degree zero and seems plausible as a unified
expression. This is the case for the RR field in the heterotic theory \cite{Fr}. A 'hybrid' 
theory like $KOSp$ might be what is needed here, in which case one could have
 that $u_2=u_1^3$. However, here we run into the same problem we had in the case of 
 K-theory: twists in low degrees will get in the way of twists in degree seven.

\vspace{3mm}
We consider the second possibility.
Let $R = \R[[v_2, v_2^{-1}]]$ be a graded ring where $v_2$ is the second generator 
of complex oriented generalized cohomology theories. It is the higher degree analog of the
Bott generator that appears in K-theory, and in general $v_n$ has dimension $2p^n-2$,
so that at the prime $p=2$, $v_2$ has dimension 6. Let $d_{H_7}=d-v_2^{-1}H_7$ 
be the twisted de Rham differential of degree one. We denote by 
$\Omega_{d_{H_7}}^i(M^{10}; \R)$ 
the space of $d_{H_7}$-closed
differential forms of total degree $i$ on $M^{10}$. 
\footnote{Subspaces of $M^{10}$ are also relevant. One such instance is if we consider the 
Hamiltonian formulation.}
In the heterotic theory we then identify the total curvature as
\(
{\mathcal F}=F_2 + v_2^{-1} F_8,
\label{gene}
\)
and thus as an element of degree two, $i=2$, in the above space of forms.
The equation $d_{H_7}{\mathcal F}=0$ defining the complex is, as shown above, just the
Bianchi identity and the equation of motion of the separate fields. The appearance of the
higher degree generator connects nicely with the discussion in \cite{KS1, KS2, KS3} 
on generalized cohomology in type II (and to some extent type I) string theories.
Furthermore, the appearance of the $w_4=0$ condition in \cite{KS1}, interpreted in 
\cite{S4} in the context of F-theory, which is a condition in heterotic string theory 
is another hint 
for the relevance of generalized cohomology in heterotic string theory. This is 
because $w_4=0$ is a condition known in type I and heterotic theory. 

\vspace{3mm}
This can be made more precise. Since we expect elliptic cohomology to also appear in 
heterotic string theory -- and in fact that is where it seems more natural as in type
IIA it was anticipated as the discussion in \cite{KS1} shows -- then we have the condition
$W_7(M^{10})=0$. This is an orientation condition in Morava K-theory and E-theory
as $W_7$ is the class corresponding to the degree seven differential in the 
Atiyah-Hirzebruch spectral sequence of these theories. Note also that the TMF 
orientation, $\frac{1}{2}p_1=0$ implies $W_7=0$ \cite{KS1}.
Given the appearance of the $H_7$-twist, we then have the modified
condition 
\(
W_7 + [H_7]=0.
\label{W7}
\)
There are two consequences. First, the analogous condition $W_3 + [H_3]=0$ amounts to 
a modification by the $H_3$-twist of the notion of ${\rm Spin}^c$ structure, which has to do with lifting the 
structure group of the tangent bundle to the connected cover. Likewise, we interpret 
$W_7 + [H_7]=0$ as a modification by the $H_7$-twist of the generalized Spin structure.
Second, we expect the condition (\ref{W7}) to correspond to a differential $d_7$ in the AHSS of 
twisted generalized theories, namely Morava K-theory and elliptic cohomology, since 
the bare differential (i.e. the one with no twist) is the first nontrivial differential in these theories, as was
shown in \cite{KS1}.  This differential is at the prime $p=2$. Hence the prime $p=2$ seems to be playing an important role in string theory. Therefore, it should not be very surprising that 
our generator in (\ref{gene}) is taken at the prime 2. Higher primes should also play 
a role, from a different point of view, and are being considered separately.

\vspace{3mm}
  We stress again that we have not settled
the question of lift to generalized cohomology, but we hope that the arguments above
would be helpful hints for achieving that goal. Some geometric aspects of the
higher twist will be discussed in \cite{SSS3}.

\section{Further Remarks}

{\bf 1. The higher twisted Chern character:} The construction of the higher-twisted Chern character 
should be interesting, possibly by a generalization of \cite{BCMMS} and \cite{AS}.
At this stage we can give the following heuristic description using the `dual' $B_6$ of the 
B-field. Let us define the twisted object
\(
{\widetilde {\cF}}=e^{\pm B_6} \cF
\)
so that
\(
d{\widetilde{\cF}}=(d \pm H_7)\wedge {\widetilde{\cF}}.
\)
This defines a twisted cohomology. More importantly, a 
higher-twisted Chern character $ch_{H_7}$, which, in the cohomologically
trivial case $H_7=dB_6$, would be of the form
\(
ch_{\pm H_7}(x)=e^{\pm B_6} ch(x).
\)

\vspace{3mm}
\noindent {\bf 2. Five-form twisting?}
One can also ask whether one could look for twisting by a five-form
field strength. The only five-form field in string theory occurs is the RR field
$F_5$ in type IIB string theory. Consider the equation
$dH_7 - F_5 \wedge H_3=0$.
If we form the differential $d_{F_5}=d- v^{-1} F_5$ of uniform degree 1, i.e.
$v$ is a generator of degree 4, then we check $d_{F_5}^2=-H_3 \wedge F_3$.
This is zero if $F_3=0$, and so we have a complex if we restrict to the NS fields.
In this case, one also has to deal with S-duality appropriately (see \cite{KS2} and 
\cite{5auth}) and this may require the use of the combined equation of motion 
\(
d(H_7 - C_0 F_7)=-F_5 \wedge(F_3 + C_0 H_3).
\)

\vspace{3mm}
\noindent {\bf 3. The Non-abelian case:}
Generalizing to the nonabelian case where 
$F_2$ is nonabelian, i.e. $F_2=d_A A$, we must replace the 
exterior derivative $d$ with the covariant exterior derivative 
$d_A$. We can see that the variation of $H_3$ will now involve the full
Chern-Simons term $CS(A)=(A\wedge dA + \frac{2}{3} A \wedge A \wedge A)$
which would give quadratic $A$ terms. However, as usual, such terms will
combine to give exactly the nonabelian field strength $F_2=d_A A$,
and so the equation in this case would then be 
$(d_A * - * H_3) \wedge F_2=0$.
The above simple manipulation of the Chern-Simons term is a direct 
consequence of the fact that both the abelian and the nonabelian 
Chern-Simons Lagrangian give the same equation of motion, $F_2=0$,
except that this is interpreted as an abelian field strength and 
a nonabelian field strength, respectively.
Here what needs to be studied is $(D-H_7)^2$. It would be interesting to 
see if an elliptic complex can be built for this case, along the lines of \cite{AHS}.

\vspace{3mm} 
\noindent {\bf 4. Relation to M-theory and other string theories:}
We would like to put the discussion of this note in the larger perspective 
of string theory and M-theory. What seems to be emerging is the following structure:
In type IIA we have the RR fields twisted by $H_3$,
in type IIB we have the NS fields twisted by $F_5$, and in the heterotic theory we 
have the gauge fields twisted by $H_7$. This motivates us to conjecture the existence 
of a grand theory in which all the notions of RR, NS and gauge fields are unified. After all,
they all come from M-theory, and in particular the M-theory $C$-field (and its `dual').
We point out that similar analysis for the M-theory fields was done in \cite{S1, S2, S3}
where it was proposed that the M-theory fields themselves should live in some 
generalized theory and attempts at characterization was made. The differential $d+G_4$ 
can be taken off-shell by taking into account the (ghost) grading of the fields and multiplying
$G_4$ by a suitable $\pm$ sign.  
We hope that the ideas in our proposals in this direction prove useful in the future.

\bigskip\bigskip
\noindent
{\bf \large Acknowledgements}

\vspace{2mm}
We thank the anonymous referee for very useful suggestions and 
 Arthur Greenspoon for comments on improving the presentation.
 
\noindent


%

\end{document}